# Modelling of Ultrawideband Propagation Scenarios for Safe Human-Robot Interaction in Warehouse Environment

Branimir Ivšić[1], Zvonimir Šipuš[1], Juraj Bartolić[1], Josip Babić[2]

[1] University of Zagreb, Faculty of electrical engineering and computing, Zagreb, Croatia,
branimir.ivsic@fer.hr, zvonimir.sipus@fer.hr, juraj.bartolic@fer.hr

[2] Končar–Electrical Engineering Institute Inc., Zagreb, Croatia,
jbabic@koncar-institut.hr

*Abstract*—The paper presents an idealized warehouse environment where human and robots are both present and need to communicate to avoid mutual collisions. The warehouse shelves are modelled as metallic (PEC) parallelepipeds, which is the largest obstacle of practical interest, while for communication an UWB Gaussian signal at 3.994 GHz and 468 MHz width is used. The signal propagation is analyzed using ray tracing software, with the goal to determine the range of communication and the optimum antenna configurations. Several scenarios have been analyzed and discussed. It is found that the major influence on the communication range is exhibited by type of polarization of antennas placed on human and robot, while the optimum range is obtained when all the antennas are vertically polarized.

## I. Introduction

The growth in logistics businesses due to internationalization of distribution chains is extensively occurring nowadays and has generated a need for larger and more evolved and efficient warehouse systems. This has motivated research in automation of warehouse operations by using robots and multi-robot systems (Fig. 1), with the goal to optimize their trajectories and avoid collisions thereby improving overall productivity and speed of warehouse management [1 – 3].

As humans still need to be present in the warehouse (e.g. for service or some intervention), safety issues arise so methods to prevent human-robot collisions are typically being implemented. To guarantee safe warehouse operation for humans the most straightforward way is to separate the robots and humans completely from one another by e.g. solid or light fences as well as laser barriers around the working area of the robots [3]. However, each time the safety barrier is crossed by a human, parts of the system or the complete system need to be shut down which increases the downtime of the system leading to higher process costs and limited operation efficiency, especially in large warehouses. As a remedy to this issue, a new integrated warehouse paradigm [4] has arisen recently where humans and robots work closely together and move freely. Alongside with the path planning algorithms [3], to ensure safety of such collaborative environment the robots are typically equipped with sensors or augmented reality technologies which enable them to avoid collisions [4]. Another layer of safety can be achieved when human and robots additionally communicate, by which the robot proximity can be evaluated, and robots can be stopped if the human is too close to them. This requires human to wear special equipment for communication such as the concept of SafetyVest [5] currently being developed. As any communication between humans and robots is to be wireless, there is also a need to explore the warehouse communication limitations on more fundamental level of electromagnetic wave propagation by which the system design limitations can be understood.

In this paper we model the typical warehouse environment with the goal to analyze the propagation of electromagnetic waves between human and robots and find the communication range that enables safe communication from the wave propagation viewpoint. We first present an idealized warehouse model and present several scenarios by which we gain an insight in the mechanisms of propagation in such environment. In the next part of the paper we study how different polarizations of the antennas placed at human and robots affect the communication range.

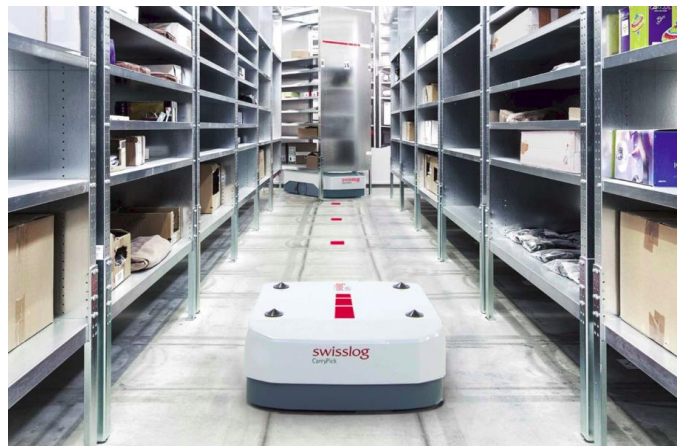

Fig. 1. The typical automated warehouse [1]

## II. The Warehouse Model

We model the warehouse shelves (racks) as parallelepipeds fully filled with metal (PEC), which represent the worst-case

This work has been supported from the European Union's Horizon 2020 research and innovation programme under grant agreement No. 688117 (SafeLog).

scenario as the propagation cannot occur through shelves. Thus, the main remaining path of propagation is through 30 cm high air gap by which the signal can be propagated merely by means of edge diffractions and multiple reflections if more shelves are present. The air gap beneath the shelves also provides path for movement of the robots (Figs 1 and 2). The antenna on the human body is placed at the height of 1.5 m (which corresponds to an average human chest height), while the antennas on the robots are placed at 20 cm height, as shown in Fig. 2.

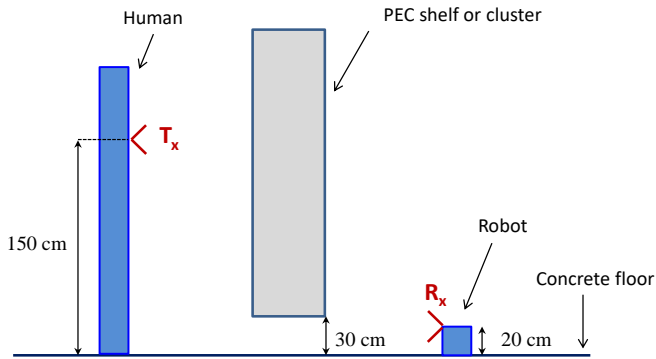

Fig. 2. The components in the proposed warehouse model

The communication between the human and robots in reality [5] is intended to be realized in UWB range (3-10 GHz) by using Decawave DW 1000 transceiver specified in [6]. Thus, in our model we establish communication via an UWB Gaussian pulse centred at 3.994 GHz and 468 MHz wide, which closely corresponds to UWB Channel 2 [6]. The antennas to be used are planar circular wideband monopoles for which several commercial prototypes have been acquired and several prototypes additionally manufactured, all of which have been shown to possess adequate dipole-like radiation characteristics in the observed range [7]. The examples of the manufactured antennas are shown in Fig. 3. For propagation model the antenna $E$-plane half-power beamwidth is taken as 60° with maximum gain of 3 dBi, which is based on the measurements presented in [7]. Furthermore, the assumed input power to the antenna is 0 dBm, while the receiver sensitivity is –106 dBm [6]. For warehouse modelling and propagation calculations we use commercially available ray-tracing software Remcom Wireless Insite [8], with full 3D propagation model and shooting and bouncing ray method.

### A. The propagation through shelves

We start the warehouse modelling with the analysis of propagation through shelves, for which a simple warehouse with only two metallic shelves (Fig. 4) is modelled. The shelves height is 3 m, while their length and thickness are 16 m and 0.5 m, respectively. The warehouse floor is made of 30 cm thick concrete ($\varepsilon_r$=7; $\sigma$=0.015 S/m), while the total floor dimensions are 40 m × 20 m. The shelves are placed symmetrically around the middle point of the warehouse and their mutual distance is 1.5 m. Both antennas on human chest (transmitting) and robot (receiving) are vertically polarized. Note that no walls are present in the warehouse as we wish to model the worst possible case and minimize the number of possible reflection paths which give rise to propagation in warehouse.

The received power distribution at 20 cm height (i.e. the robot positions) along the warehouse with two shelves is shown in Figs 5 and 6, for the cases when the transmitting antenna is placed between the shelves and when the transmitting antenna is in free half-space around 1m close to the first shelf, respectively. It can be seen that better signal coverage across warehouse is obtained when the human (i.e. transmitting antenna) is placed between shelves (Fig. 5) as in that case there is mechanism for multiple reflections by which the energy is propagated through the air gaps below shelves (i.e. the space between the shelves acts as a lossy resonator). Analogous observations occur when number of shelves is increased (as illustrated e.g. for the case of 16 shelves shown in Figs 7 and 8) meaning that for good communication it is important to have some mechanism of multiple reflections near the transmitting antenna by which further communication via air gaps is supported regardless of the number of shelves.

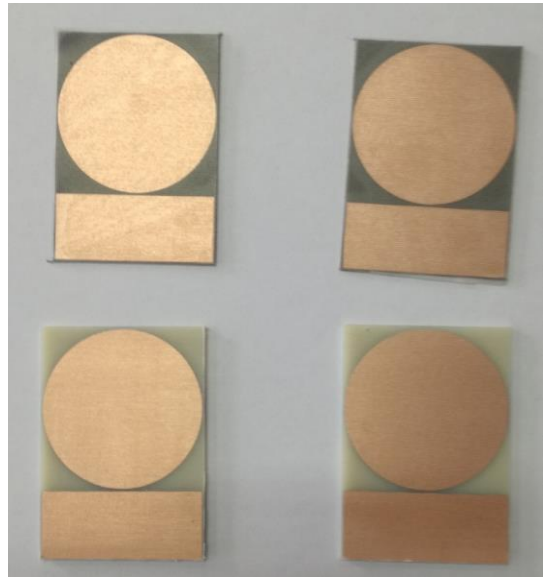

Fig. 3. Examples of manufactured UWB planar circular monopole antennas for human-robot communication (circle diameter: 3 cm). Upper pair: Taconic substrate; lower pair: FR408 HR substrate [7].

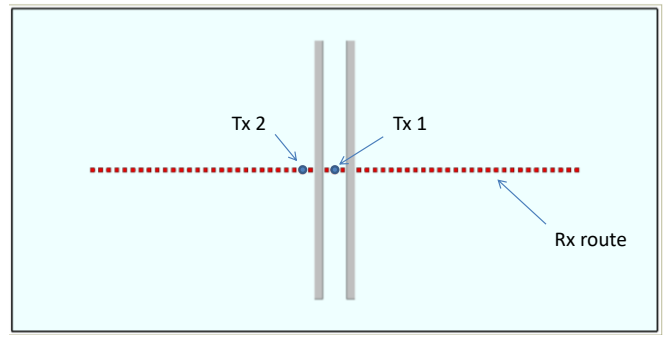

Fig. 4. The layout of basic warehouse model with two metallic shelves (top view)

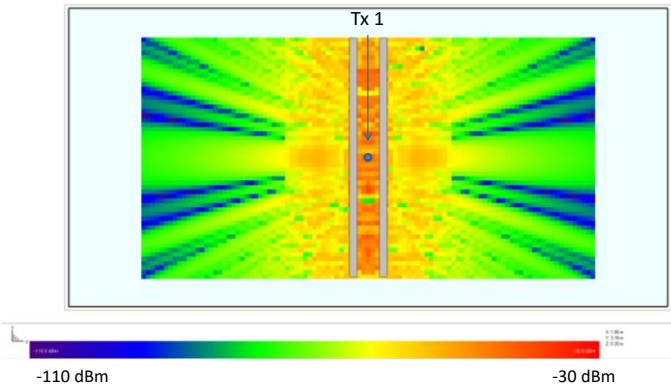

-110 dBm          -30 dBm

Fig. 5. Received power when the transmitter is placed in the middle of the warehouse (case with 2 shelves)

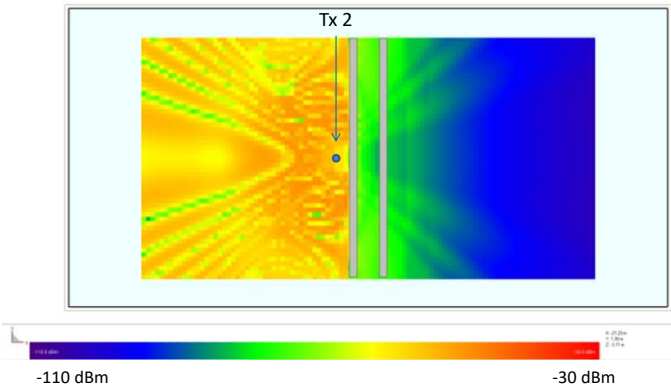

-110 dBm          -30 dBm

Fig. 6. Received power when the transmitter is placed before leftmost shelf (case with 2 shelves)

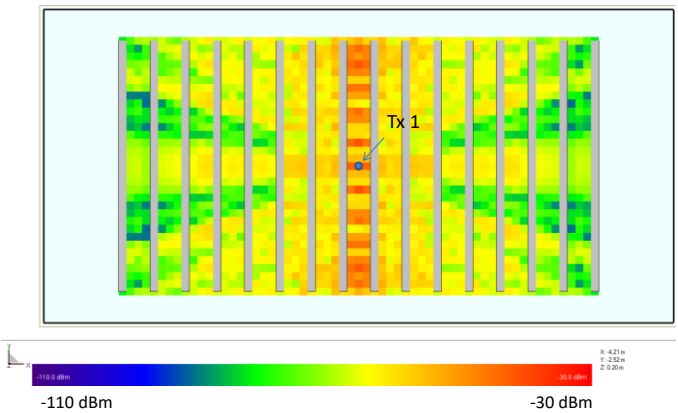

-110 dBm          -30 dBm

Fig. 7. Received power when the transmitter is placed in the middle of the warehouse (case with 16 shelves)

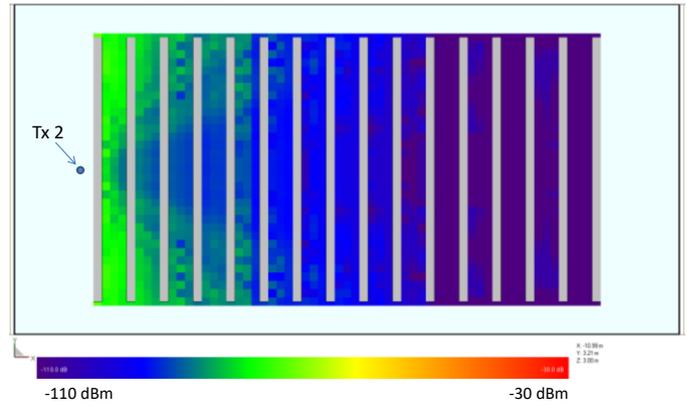

-110 dBm          -30 dBm

Fig. 8. Received power when the transmitter is placed before leftmost shelf (case with 16 shelves)

### B. The refined warehouse model

We now proceed to the more realistic case where shelves (still made of full metal) are of dimensions suitable to be carried by robots. The height of new shelves is 2 m, while its horizontal dimensions are 1.3 m × 1.3 m. The shelves are grouped in clusters of 7×2 shelves (Fig. 9), while between clusters the 1.5 m wide corridor is placed. In addition, in refined model we allow small 5 cm gap between shelves in clusters which also enhances communication. This model is considered as the worst realistic case (and is somewhat relaxed compared to the case in previous section).

Figure 10 shows the received power profile within the shelf area of the warehouse when the transmitter (i.e. the human) is placed in the middle of the warehouse, while Fig. 11 shows the received power profile when the transmitter is placed in free space 2 meters before shelf cluster (the shelf clusters are shown in wireframe mode for convenience and clear picture). In both cases adequate signal coverage is obtained (vertical polarization of all the antennas is assumed here). For the case of Fig. 10 it can also be noted that the maximum signal is actually not directly under the transmitter, which is expected as this area is in the null of the supposed dipole radiation pattern, meaning that the direct signal component there is low.

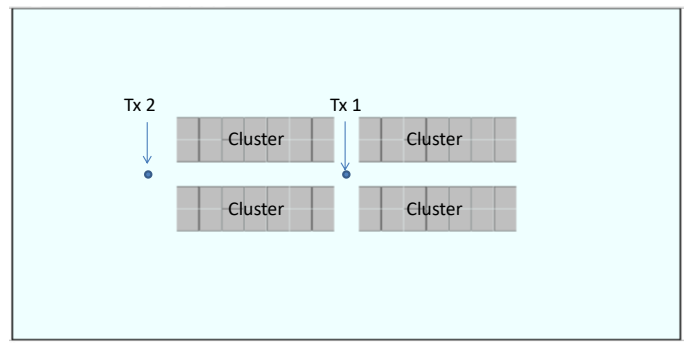

Fig. 9. The layout of the refined warehouse with four 7×2 shelf clusters

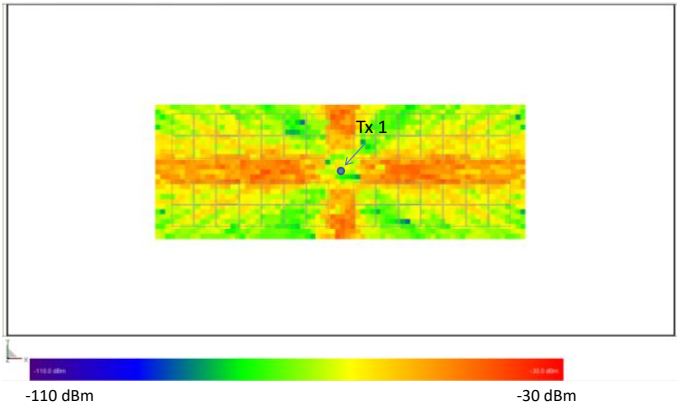

-110 dBm                                      -30 dBm

Fig. 10. Received power when the transmitter is placed in the middle of the warehouse (case with four shelf clusters)

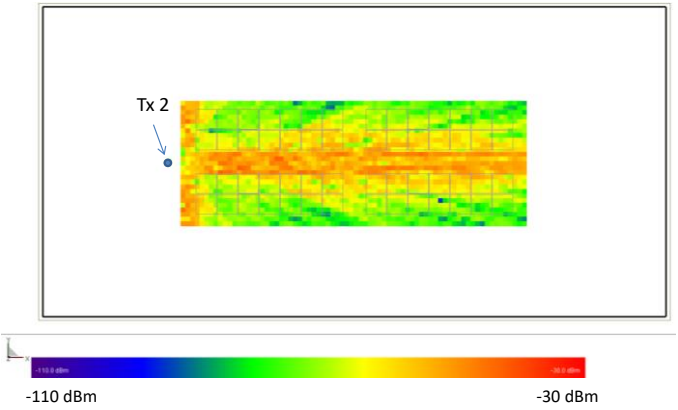

-110 dBm                                      -30 dBm

Fig. 11. Received power when the transmitter is placed before leftmost shelf cluster (case with four shelf clusters)

## III. OTHER POLARIZATION SETUPS

From the previous section it can be inferred that in the realistic warehouse even for worst case scenario it is possible to establish safe communication to over 20 meters of distance regardless whether human is placed in the middle or at the end of the warehouse. This possibility is nevertheless by now shown to occur when all the antennas are vertically polarized. Thus, the next point of interest is to discuss other possible polarization configurations in order to find out how much the antenna polarization affects propagation and the communication range. In Figs 12 – 14 the main results are shown for various relative orientations of transmitting and receiving antennas. By comparing Figs 12 and 14 it can be noted that when receiving antennas are horizontally polarized the signal coverage is worse compared to the case when receiving antenna is vertically polarized. This is also seen by comparing Figs 12 and 11, where for horizontal polarization more "blind spots" occur.

This difference in polarization response can be explained by noting that the signal from the vertically polarized antenna predominantly exhibits parallel ($H$) polarization [9] with respect to the ground where the electric field has a significant vertical component which propagates under the shelves better than its horizontal counterpart, where due to boundary conditions an image field from the metallic shelf cancels it. On the other hand, the electric field from horizontally polarized antenna impinges on the ground predominantly in perpendicular ($E$-) polarization where the horizontal component of electric field largely dominates. This point however needs further verification in the future as the propagation mechanism in warehouse environment is rather complex to predict in deterministic terms.

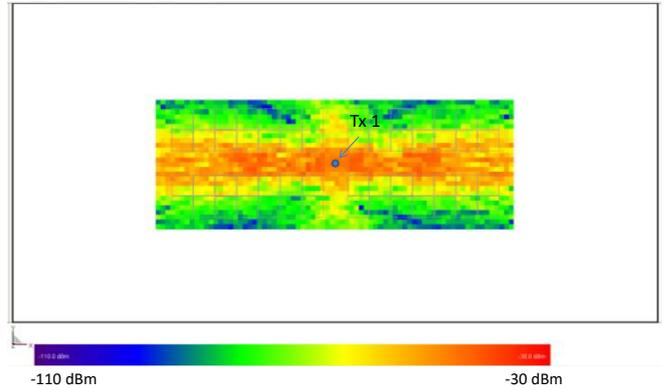

-110 dBm                                      -30 dBm

Fig. 12. Received power when both transmitter and receiver are horizontally polarized (null in transverse direction)

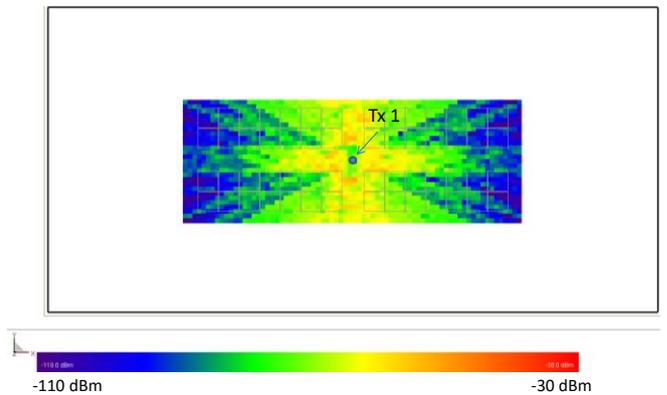

-110 dBm                                      -30 dBm

Fig. 13. Received power when transmitter is horizontally polarized (null in transverse direction) while receiver is vertically polarized

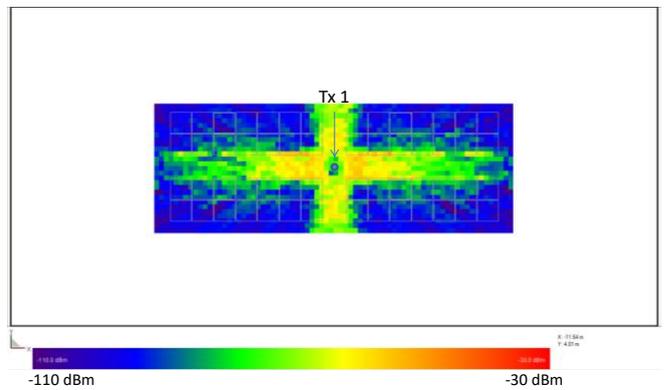

-110 dBm                                      -30 dBm

Fig. 14. Received power when transmitter is vertically polarized, while receiver is horizontally polarized (null in transverse direction)

## A. The "lying man" isssue

Another problem that arises in practice is that the human can change its body position, while in warehouse (he can e.g. kneel or even lie down on the floor if some warehouse task or service requires it or if there is some accident). This of course changes the orientation (and hence polarization) of the transmitting antenna on the human body. To analyse this scenario the transmitter is hence assumed to be at height of 0.2 m which is the same height as robots (i.e. it represents the case of lying human). The three possible transmitting antenna polarizations are analyzed in Figs 15 – 17 (the antenna on robot is assumed fixed with vertical polarization). It can be seen that an optimum signal coverage occurs when transmitting antenna is vertically polarized (which is expected), while for other orientations the transmitted power is significantly lower and can result in signal loss. This issue can be mitigated e.g. by mounting two antennas with different polarizations on the human body (i.e. using polarization diversity) and is to be considered in our future work.

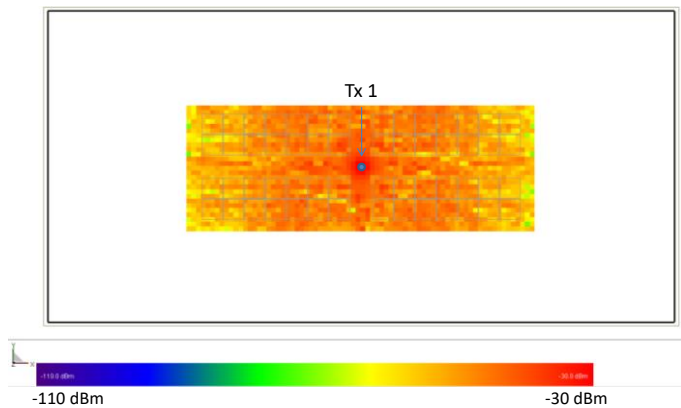

Fig. 15. Received power when both transmitter (at 0.2 m height) and receiver are vertically polarized

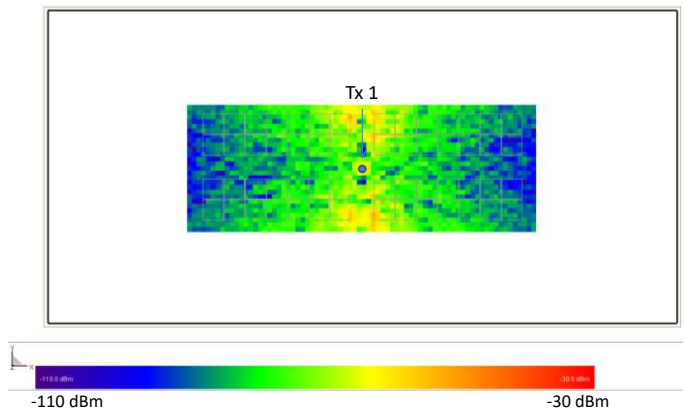

Fig. 16. Received power when the receiver is vertically polarized, while transmitter (at 0.2 m height) is horizontally polarized (null in longitudinal direction)

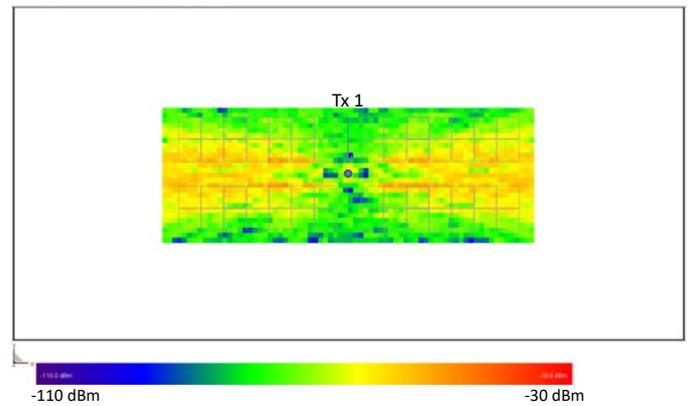

Fig. 17. Received power when the receiver is vertically polarized, while transmitter (at 0.2 m height) is horizontally polarized (null in transverse direction)

## IV. CONCLUSION

In this paper we have presented the model of the typical collaborative warehouse environment where both humans and robots are present and need to communicate safely to avoid collisions. We have analysed the propagation of electromagnetic waves using ray-tracing method with the goal to find the communication range that enables safe communication for several scenarios of interest. That way we have obtained fundamental limitations on communication range in warehouse environment, which can further serve as a guideline in logistic system design and estimation of the possible safety level of communication. In first warehouse model we have found that multiple reflections and diffractions are the dominant mechanism of propagation through the air gap under the shelves and that the space between shelves acts as a resonator which couples the energy below shelves, and that when the resonator effect is present it is possible to establish safe communication to over 20 m distance. In the second, more realistic model of warehouse, different polarizations of the antennas have been analyzed and it was found that optimum communication is achieved when all the antennas are vertically polarized, while some "blind spots" occur when any of the antennas in configuration is horizontally polarized.

The future work will consist of improving the warehouse propagation model by evaluation of more cases of interest which might occur in reality, such as an impact of rough surfaces, antenna height variations, influence of the human body etc. In addition, the simple model to explain the dominant mechanisms of propagation for vertical and horizontal polarization is to be obtained.